\PassOptionsToPackage{unicode}{hyperref}
\PassOptionsToPackage{hyphens}{url}
\PassOptionsToPackage{dvipsnames,svgnames,x11names}{xcolor}
\documentclass[
  letterpaper,
  DIV=11,
  numbers=noendperiod]{scrartcl}
\usepackage{xcolor}
\usepackage{amsmath,amssymb}
\usepackage{iftex}
\ifPDFTeX
  \usepackage[T1]{fontenc}
  \usepackage[utf8]{inputenc}
  \usepackage{textcomp} 
\else 
  \usepackage{unicode-math} 
  \defaultfontfeatures{Scale=MatchLowercase}
  \defaultfontfeatures[\rmfamily]{Ligatures=TeX,Scale=1}
\fi
\usepackage{lmodern}
\ifPDFTeX\else
\fi
\IfFileExists{upquote.sty}{\usepackage{upquote}}{}
\IfFileExists{microtype.sty}{
  \usepackage[]{microtype}
  \UseMicrotypeSet[protrusion]{basicmath} 
}{}
\makeatletter
\@ifundefined{KOMAClassName}{
  \IfFileExists{parskip.sty}{%
    \usepackage{parskip}
  }{
    \setlength{\parindent}{0pt}
    \setlength{\parskip}{6pt plus 2pt minus 1pt}}
}{
  \KOMAoptions{parskip=half}}
\makeatother
\makeatletter
\ifx\paragraph\undefined\else
  \let\oldparagraph\paragraph
  \renewcommand{\paragraph}{
    \@ifstar
      \xxxParagraphStar
      \xxxParagraphNoStar
  }
  \newcommand{\xxxParagraphStar}[1]{\oldparagraph*{#1}\mbox{}}
  \newcommand{\xxxParagraphNoStar}[1]{\oldparagraph{#1}\mbox{}}
\fi
\ifx\subparagraph\undefined\else
  \let\oldsubparagraph\subparagraph
  \renewcommand{\subparagraph}{
    \@ifstar
      \xxxSubParagraphStar
      \xxxSubParagraphNoStar
  }
  \newcommand{\xxxSubParagraphStar}[1]{\oldsubparagraph*{#1}\mbox{}}
  \newcommand{\xxxSubParagraphNoStar}[1]{\oldsubparagraph{#1}\mbox{}}
\fi
\makeatother

\usepackage{longtable,booktabs,array}
\usepackage{calc} 
\usepackage{etoolbox}
\makeatletter
\patchcmd\longtable{\par}{\if@noskipsec\mbox{}\fi\par}{}{}
\makeatother
\IfFileExists{footnotehyper.sty}{\usepackage{footnotehyper}}{\usepackage{footnote}}
\makesavenoteenv{longtable}
\usepackage{graphicx}
\makeatletter
\newsavebox\pandoc@box
\newcommand*\pandocbounded[1]{
  \sbox\pandoc@box{#1}%
  \Gscale@div\@tempa{\textheight}{\dimexpr\ht\pandoc@box+\dp\pandoc@box\relax}%
  \Gscale@div\@tempb{\linewidth}{\wd\pandoc@box}%
  \ifdim\@tempb\p@<\@tempa\p@\let\@tempa\@tempb\fi
  \ifdim\@tempa\p@<\p@\scalebox{\@tempa}{\usebox\pandoc@box}%
  \else\usebox{\pandoc@box}%
  \fi%
}
\def\fps@figure{htbp}
\makeatother

\NewDocumentCommand\citeproctext{}{}

\makeatletter
 \let\@cite@ofmt\@firstofone
 \def\@biblabel#1{}
 \def\@cite#1#2{{#1\if@tempswa , #2\fi}}
\makeatother
\newlength{\cslhangindent}
\newlength{\csllabelwidth}
\newenvironment{CSLReferences}[2] 
 {\begin{list}{}{%
  \setlength{\itemindent}{0pt}
  \setlength{\leftmargin}{0pt}
  \setlength{\parsep}{0pt}
  \ifodd #1
   \setlength{\leftmargin}{\cslhangindent}
   \setlength{\itemindent}{-1\cslhangindent}
  \fi
  \setlength{\itemsep}{#2\baselineskip}}}
 {\end{list}}
\usepackage{calc}

\providecommand{\tightlist}{%
  \setlength{\itemsep}{0pt}\setlength{\parskip}{0pt}}

\usepackage{booktabs}
\usepackage{caption}
\usepackage{longtable}
\usepackage{colortbl}
\usepackage{array}
\usepackage{anyfontsize}
\usepackage{multirow}
\KOMAoption{captions}{tablesignature}
\makeatletter
\@ifpackageloaded{caption}{}{\usepackage{caption}}
\AtBeginDocument{%
\ifdefined\contentsname
  \renewcommand*\contentsname{Table of contents}
\else
  \newcommand\contentsname{Table of contents}
\fi
\ifdefined\listfigurename
  \renewcommand*\listfigurename{List of Figures}
\else
  \newcommand\listfigurename{List of Figures}
\fi
\ifdefined\listtablename
  \renewcommand*\listtablename{List of Tables}
\else
  \newcommand\listtablename{List of Tables}
\fi
\ifdefined\figurename
  \renewcommand*\figurename{Figure}
\else
  \newcommand\figurename{Figure}
\fi
\ifdefined\tablename
  \renewcommand*\tablename{Table}
\else
  \newcommand\tablename{Table}
\fi
}
\@ifpackageloaded{float}{}{\usepackage{float}}
\floatstyle{ruled}
\@ifundefined{c@chapter}{\newfloat{codelisting}{h}{lop}}{\newfloat{codelisting}{h}{lop}[chapter]}
\floatname{codelisting}{Listing}

\makeatother
\makeatletter
\@ifpackageloaded{caption}{}{\usepackage{caption}}
\@ifpackageloaded{subcaption}{}{\usepackage{subcaption}}
\makeatother
\usepackage{bookmark}
\IfFileExists{xurl.sty}{\usepackage{xurl}}{} 
\hypersetup{
  pdftitle={Correcting Mode Collapse in Silicon Sampling with Semantic Similarity Rating},
  pdfauthor={Oscar Heath; Rohan Alexander},
  colorlinks=true,
  linkcolor={blue},
  filecolor={Maroon},
  citecolor={Blue},
  urlcolor={Blue},
  pdfcreator={LaTeX via pandoc}}

\title{Correcting Mode Collapse in Silicon Sampling with Semantic
Similarity Rating\thanks{Code and data are available at:
\url{https://doi.org/10.5281/zenodo.21706514}. We gratefully acknowledge
funding from Natural Sciences and Engineering Research Council of Canada
(NSERC) Alliance Grant (110\_2024\_2025\_Q4\_13) and Social Sciences and
Humanities Research Council (SSHRC) Partnership Grant (895-2025-1002).
This research was conducted while Oscar Heath was an intern at the
Investigative Journalism Foundation. We thank Benedict Cummins-Mburu,
Ellie Murray, Mariana Garcia Mejia, Sabrina Kreyzerman, and Zane
Schwartz, for comments and suggestions. Additionally, we thank Annie
Collins, Ricardo Baptista, and Chagai Weiss for their comments on early
drafts of the paper. Correspondence:
\href{mailto:rohan.alexander@utoronto.ca}{\nolinkurl{rohan.alexander@utoronto.ca}}.}}
\author{Oscar Heath \and Rohan Alexander}
\date{July 30, 2026}
\begin{document}
\maketitle
\begin{abstract}
Silicon sampling refers to the use of Large Language Models (LLMs) to
generate responses to surveys. It has shown promise, but tends to
generate response distributions with unrealistically low variance. We
argue that this mode collapse is due to LLMs failure to generate numeric
data, and that text responses may be better suited for this task. We
analyze whether Semantic Similarity Rating can improve the fidelity of
silicon sampling responses when asked about political attitudes. This
method solicits text-only responses from LLMs, then maps this to a
numeric scale using text embeddings. We find that this method both
improves the fidelity of silicon sampling response distributions, and
has few parameters to calibrate.\\

\end{abstract}

\section{Introduction}\label{introduction}

In 2013, the National Academy of Sciences reported declining rates over
the two previous decades (National Research Council 2013). Response
rates to household surveys have been consistently declining, with the
U.S. Current Population Survey reaching historically low response rates
in November 2025 (Brookings 2026). As response rates decline, the
effectiveness of random sampling has increasingly come into question,
and researchers have turned to new methods such as sophisticated
re-weighting approaches, opt-in surveys, and online panels.

Researchers have recently found that LLMs can be conditioned via prompts
to simulate human survey responses, which can be quite accurate. This
process, coined ``Silicon Sampling'' by Argyle et al. (2023), allows
data to be generated for a trivial cost in a fraction of the time
compared with obtaining responses from human participants. This emergent
property of LLMs has been studied in various fields like political
science (Argyle et al. 2023), psychology (Dillion et al. 2023), and
economics (Horton et al. 2026). In general, their findings show that LLM
responses are very similar to human responses. In the private sector,
companies are beginning to make use of synthetic customers for market
research, and companies selling synthetic data and personas are emerging
(The New York Times 2026a; Bain and Company 2026).

Despite the growing adoption of silicon sampling, the degree to which it
can be trusted remains unclear. One concern that has arisen in multiple
studies is the prevalence of mode collapse, a phenomenon where the model
generates responses that are overly concentrated around the mode of the
true distribution. This causes misleading results: synthetic means are
quite close to the true means, but analysing the entire distribution
shows unrealistically low variance (Bisbee et al. 2024; Kaiser et al.
2025; Barrie and Cerina 2026). This significantly impacts downstream
analyses, leading to overconfident estimates, narrow confidence
intervals, and incorrect power calculations. Additionally, silicon
samples are affected by minor prompt changes, as well as choice and
version of model (Alexander and Collins 2026; Schröder et al. 2025;
Bisbee et al. 2024).

In this paper, we revisit the analysis of Bisbee et al. (2024), aiming
to correct mode collapse. We generate samples of thermometer scores, a
0-100 survey measure of respondent's feelings on a given topic. We
compare two methods of generating silicon samples. In the first, we
directly prompt the model for numeric output --- the approach used by
Bisbee et al. (2024), as well as many other silicon sampling studies. In
the second, we apply Semantic Similarity Rating (SSR) (Maier et al.
2025): instead of prompting LLMs for numeric scores, we generate purely
text responses, then map these responses to a thermometer scale using
text embeddings. We hypothesize that this method is better suited for
silicon sampling, as LLMs are predominantly trained to generate text. We
measure the fidelity of the synthetic distributions using the
Kullback-Leibler (KL) divergence compared to the true distribution,
which is a measure of how well the synthetic distribution approximates
the real distribution. Additionally, we compare the absolute error of
synthetic means to the true means, by respondent group and target group.

We find that applying SSR to text outputs produces synthetic response
distributions that are much closer to real ANES (American National
Election Studies 2016) distributions, with lower KL divergence, and no
significant loss in accuracy of the synthetic means. The issue of low
variance seems effectively addressed by applying SSR with a global
temperature parameter. This is a single learned parameter that controls
the variance of the generated distributions. We found no noticeable
degradation in performance when applying the temperature parameter
learned on 2016 ANES data to 2020 ANES data, suggesting that the
parameter generalizes well to future data. However, some limitations
remain: while SSR improves the variance of synthetic response
distributions the means of synthetic distributions (both from numeric
responses and SSR) are still consistently biased, in some cases,
relative to real data, just as Bisbee et al. (2024) found.

The remainder of the paper is structured as follows: In
Section~\ref{sec-background}, we review related work in silicon sampling
and machine learning calibration, and situate our contributions to the
field. In Section~\ref{sec-methods}, we describe our experiment design,
including how we generate silicon samples, apply SSR, and calibrate
temperature parameters. In Section~\ref{sec-results}, we present the
results of our analysis. Finally, in Section~\ref{sec-discussion}, we
discuss the implications of our findings, limitations of our study, and
directions for future research.

\section{Background}\label{sec-background}

LLMs have been widely applied in political science, with encouraging
results. For instance, Le Mens and Gallego (2025) use LLMs for
positioning political texts along various ideological axes, which
achieves over 0.9 correlation with human expert ratings. Benoit et al.
(2026) use LLMs to extract policy positions from political manifestos
and long texts, also correlating highly with expert ratings. Argyle et
al. (2023) found that LLMs prompted with personas generate mean
responses to American National Election Survey (ANES) questions that are
very close to true responses. Motivated by these findings, researchers
in the field of silicon sampling have experimented with using LLMs to
generate synthetic survey responses. Park et al. (2026) even proposes a
framework for LLMs as ``general-purpose simulation of individuals''
(Park et al. 2026, 1).

In the midst of this growing body of research in silicon sampling, many
researchers have also pointed out shortcomings of LLM-generated data.
Our research builds on Bisbee et al. (2024), whose critical analysis of
silicon sampling found that LLMs generate synthetic responses with
unrealistically low variance compared to real data. This prohibits
researchers from performing downstream quantitative analyses on silicon
data. Bisbee et al. (2024) uses OpenAI's ChatGPT 3.5 Turbo and ChatGPT
4.0 out-of-the-box, directly prompting the model to generate numeric
responses to thermometer questions. We argue that this method of
prompting may not be optimal for generating synthetic responses that
reflect the underlying distribution of real responses, and that
alternative methods of prompting and post-processing may improve the
fidelity of silicon sampling.

Fundamentally, silicon sampling aims to generate responses that reflect
the underlying distribution of real responses. Theoretically, this is
made possible by the fact that LLMs are trained on such large amounts of
data that they have learned these distributions. Yet, even when given
perfect information about a statistical distribution, for instance, a
standard Normal distribution, LLMs are not able to accurately generate
random samples from this distribution without the use of external tools
(Zhao et al. 2026).

Hence, we hypothesize that common failures of LLMs in silicon sampling,
such as low variance, may not be due to a misunderstanding of underlying
response distributions, but rather due to an inability to generate
samples that reflect said distribution. Fundamentally, LLMs are
probabilistic models trained to predict the next token in a sequence,
and excel at generating natural language text. Instead of prompting the
model to directly generate a numeric response, we employ Semantic
Similarity Rating (SSR), a method that solicits text-only responses from
LLMs, then maps this output to a probability distribution over a 0-100
scale using text embeddings and cosine similarity to predefined anchor
points on that scale (Maier et al. 2025). This approach has two
benefits: first, it plays to one of the main strengths of LLMs, which is
generating natural language text, rather than numeric responses. Second,
it allows us to generate a full probability distribution over the
response scale for each sample, rather than a single point estimate. SSR
has been used for simulating buyer intent and ratings (Maier et al.
2025; Pichardo 2026), outperforming other methods of generating numeric
responses. In this paper, we systematically study its application to
silicon sampling for political science research.

Other methods of improving algorithmic fidelity have been proposed and
also present promising avenues of research. Cao et al. (2025) and Suh et
al. (2026) use fine-tuning to train LLMs on past data, finding strong
performance gains. However, this approach is constraining in multiple
ways: it requires access to model weights (which is often impossible for
frontier LLMs), large amounts of training data, and significant
computational resources. Chapala et al. (2025) found that
``psychologically grounded prompt wording'' can mitigate social
desirability bias in silicon sampling, and pull LLM response
distributions closer to real responses on the ANES survey. Nevertheless,
this approach is limited in that it requires manual prompt engineering,
and Chapala et al. (2025) found inconsistent results among different
models and sample questions. In contrast, SSR offers a simple
training-free approach to improving fidelity, requiring less manual
prompt engineering than Chapala et al. (2025), or access to model
weights.

\section{Methods}\label{sec-methods}

We chose to make silicon samples of thermometer scores, a common survey
instrument where respondents are asked to rate their feelings toward a
certain target group between 0 and 100. We generate synthetic
thermometer scores for respondents of the 2016 ANES survey, for four
target groups: Democratic Party, Republican Party, Liberals, and
Conservatives. For each real respondent, we create a persona prompt,
which is a text description of the respondent's demographic and
political characteristics, which is passed into various frontier LLMs as
a prompt. For each persona, we generate two types of responses:

\begin{enumerate}
\def\labelenumi{\arabic{enumi}.}
\item
  \textbf{Numeric response}: we directly prompt the model to generate a
  numeric response between 0 and 100.
\item
  \textbf{SSR response}: we prompt the model to generate a text response
  describing its feelings toward the target group. We then map this text
  response to a probability distribution over the thermometer scale
  using SSR.
\end{enumerate}

The SSR process has a single learned parameter \(\mathcal{T}\), which
controls the variance of the generated distribution. We learn this
parameter by choosing the value that minimizes the Kullback-Leibler (KL)
divergence between the real and synthetic distributions for 2016 ANES
data across all groups and questions. KL divergence measures how well
the synthetic distribution approximates the real distribution, with
lower values indicating a better match. We then apply this parameter to
2020 ANES data, to test whether it generalizes to future data.

\begin{figure}

\centering{

\pandocbounded{\includegraphics[keepaspectratio]{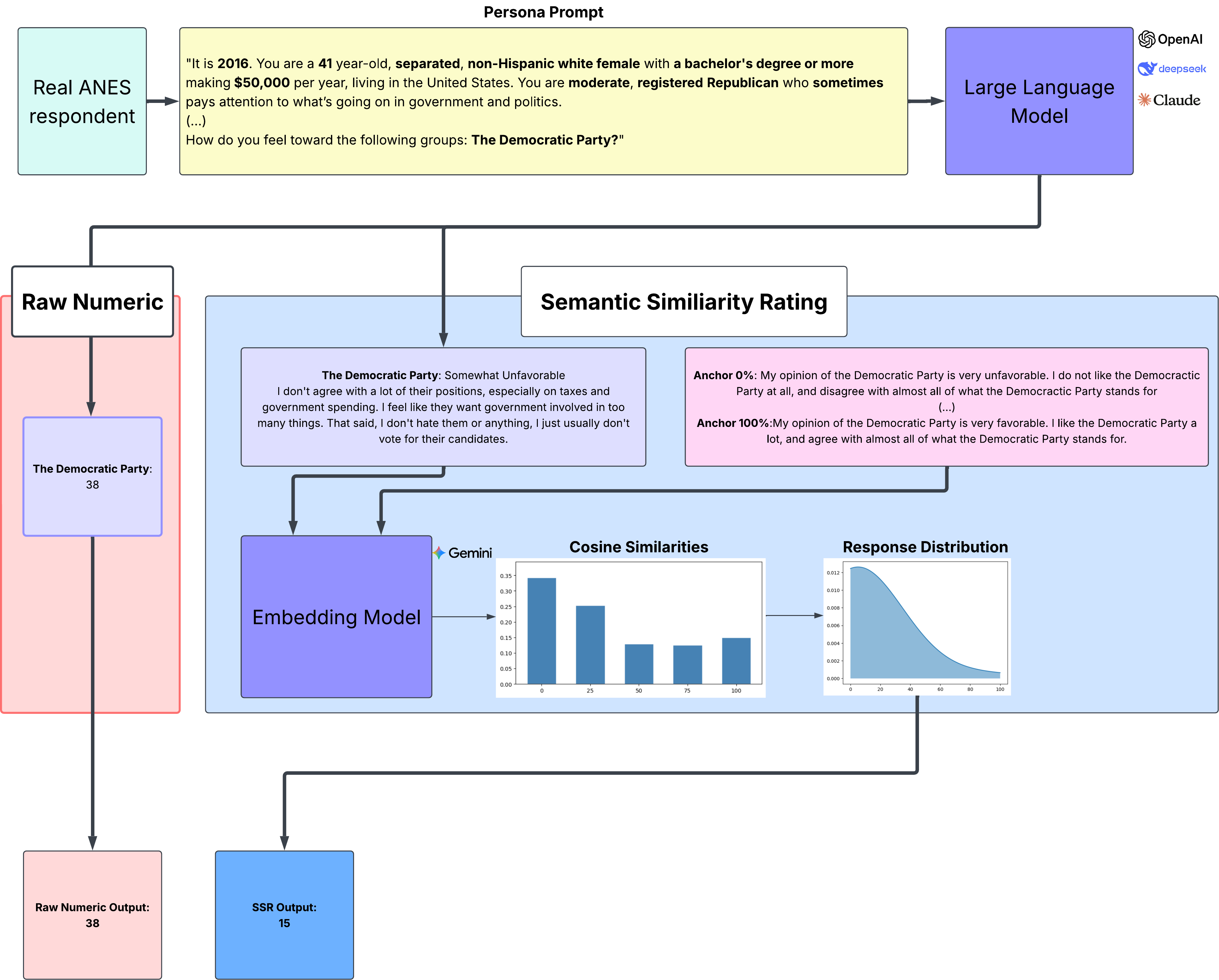}}

}

\caption{\label{fig-ssr}Overview of numeric and SSR silicon sampling
processes}

\end{figure}

A high level overview of the numeric and SSR silicon sampling processes
is shown in Figure~\ref{fig-ssr}.

\subsection{Data}\label{data}

We use data from the reproducibility package of Bisbee et al. (2024),
which contains their samples from the 2016 ANES Time Series Study
(American National Election Studies 2016) and their prompts.
Additionally, we use the 2020 ANES Time Series Study as a test set
(American National Election Studies 2020).

\subsection{Prompting}\label{prompting}

To generate personas, we used the same prompts as Bisbee et al. (2024),
which have the following format:

``It is {[}\texttt{YEAR}{]}. You are a {[}\texttt{AGE}{]} year-old,
{[}\texttt{MARST}{]}, {[}\texttt{RACETH}{]} {[}\texttt{GENDER}{]} with
{[}\texttt{EDUCATION}{]} making {[}\texttt{INCOME}{]} per year, living
in the United States. You are {[}\texttt{IDEO}{]}, {[}\texttt{REGIS}{]}
{[}\texttt{PID}{]} who {[}\texttt{INTEREST}{]} pays attention to what's
going on in government and politics.''

Terms in square brackets are replaced with the corresponding information
from the ANES respondents (see Appendix~\ref{sec-prompts} for variable
information). We used two variants of the prompt, one soliciting a
numeric response, and one asking the model to generate only a text
response to the thermometer question.

We generated responses using four models:
\texttt{deepseek-v4-flash}(DeepSeek-AI 2026), \texttt{claude-sonnet-5}
(Anthropic 2026), \texttt{gpt-5.4-mini} (OpenAI 2026). We generated one
response per persona, for a total of 7,530 responses per prompt variant.
We asked synthetic respondents for their feelings toward the Democratic
Party, Republican Party, Liberals, and Conservatives, for a total of
four questions per respondent. For the numeric prompt, we directly
prompted the model to generate a numeric response between 0 and 100. For
the text prompt, we asked the model to generate a text response
describing their feelings toward the target group, without any mention
of a numeric scale. The prompts used for each model are as follows:

Numeric prompt:

Responses should be given on a scale from 0 (meaning cold feelings) to
100 (meaning warm feelings). Ratings between 50 degrees and 100 degrees
mean that you feel favorable and warm toward the group. Ratings between
0 degrees and 50 degrees mean that you don't feel favorable toward the
group and that you don't care too much for that group. You would rate
the group at the 50 degree mark if you don't feel particularly warm or
cold toward the group.

Text prompt:

The following questions ask about individuals' feelings toward different
groups. Responses should be your genuine feelings toward each group, not
what you think is socially desirable. Please elaborate in your answers.

\subsection{Semantic Similarity
Rating}\label{semantic-similarity-rating}

Figure~\ref{fig-ssr} outlines the SSR pipeline, which maps text
responses to a probability distribution over the thermometer scale. We
pre-determine five anchor points on the thermometer scale, to which we
assign text descriptions corresponding to their relative magnitude. For
instance, the anchor for the score of 0 is ``I have a very unfavorable
opinion of {[}GROUP{]}. I strongly dislike {[}GROUP{]} and reject almost
everything {[}GROUP{]} stands for.'' (see
Appendix~\ref{sec-anchor-points} for the full list of anchor points). We
then generate text embeddings of all anchors, as well as for each
synthetic respondents' text response to the thermometer question. Next,
we calculate cosine similarities between the response and the anchors,
passing these through normalization, temperature-scaled softmax, and
kernel density estimation to generate a smooth probability distribution
over the thermometer scale for each response.

Given text responses and anchor points, we embed them using
\texttt{Gemini\ Embedding\ 2} (Shanbhogue et al. 2026), which generates
a dense vector representation of each text. Following Pichardo (2026),
we use asymmetric embedding: we frame this as a document retrieval
problem, embedding the anchors as documents, and the response as a
query, which empirically improves downstream similarity ratings. We
choose to use \texttt{Gemini\ Embedding\ 2} for its strong performance
on the MTEB text-to-text embedding benchmark (Muennighoff et al. 2023;
Shanbhogue et al. 2026), relative price point, and support for
asymmetric embedding.

We then calculate the cosine similarity between the response embedding
\(E_{response}\) and embedded anchor points \(E_{anchor_i}\) on the
thermometer scale (See Appendix~\ref{sec-anchor-points} for details),
generating a list of similarities \(S = \{s_1, s_2, s_3, s_4, s_5\}\)
(see Equation~\ref{eq-cosine-similarity}).

\begin{equation}\protect\phantomsection\label{eq-cosine-similarity}{
s_i = \frac{E_{response} \cdot E_{anchor_i}}{\|E_{response}\| \|E_{anchor_i}\|}
}\end{equation}

Because the global embedding space allows for representation of a wide
variety of text, the embeddings of all ANES responses end up lying very
close together (\textasciitilde0.1 distance). Hence, we apply
normalization to the cosine similarities to increase the spread of the
generated distributions, following Pichardo (2026). This is done by
first calculating the minimum and maximum similarity scores across all
anchors for a given response, and then applying min-max normalization to
scale the similarities to a 0-1 range (Equation~\ref{eq-normalization}).

\begin{equation}\protect\phantomsection\label{eq-normalization}{
s_{i, normalized} = \frac{s_i - \min(S)}{\max(S) - \min(S)}
}\end{equation}

After normalization, we apply a softmax function to the normalized
similarities to scale similarities into a probability mass function,
with an additional temperature parameter \(\mathcal{T}\) to control the
variance of the distribution (Equation~\ref{eq-softmax}).

We empirically generated distributions for all values of \(\mathcal{T}\)
between 0 and 1, at increments of 0.05, finding that
\(\mathcal{T} = 0.25\) minimized the mean KL divergence between the real
and synthetic distributions for 2016 ANES data across all groups and
questions. KL divergence is a measure of how well the synthetic
distribution approximates the real distribution, with lower values
indicating a better match. We use this temperature parameter for all
subsequent analyses.
\begin{equation}\protect\phantomsection\label{eq-softmax}{
s_{i, softmax} = \frac{\exp(s_{i, normalized} / \mathcal{T})}{\sum_{j=1}^n \exp(s_{j, normalized} / \mathcal{T})}
}\end{equation}

Finally, given the five softmax-scaled similarities, we generate a
probability distribution over the thermometer scale. We treat the
softmax-scaled similarities as weights for each anchor point on the
thermometer scale, and generate a Gaussian kernel density estimate to
produce a smooth probability distribution \(Q_t\) over the thermometer
scale.

\begin{figure}

\centering{

\pandocbounded{\includegraphics[keepaspectratio]{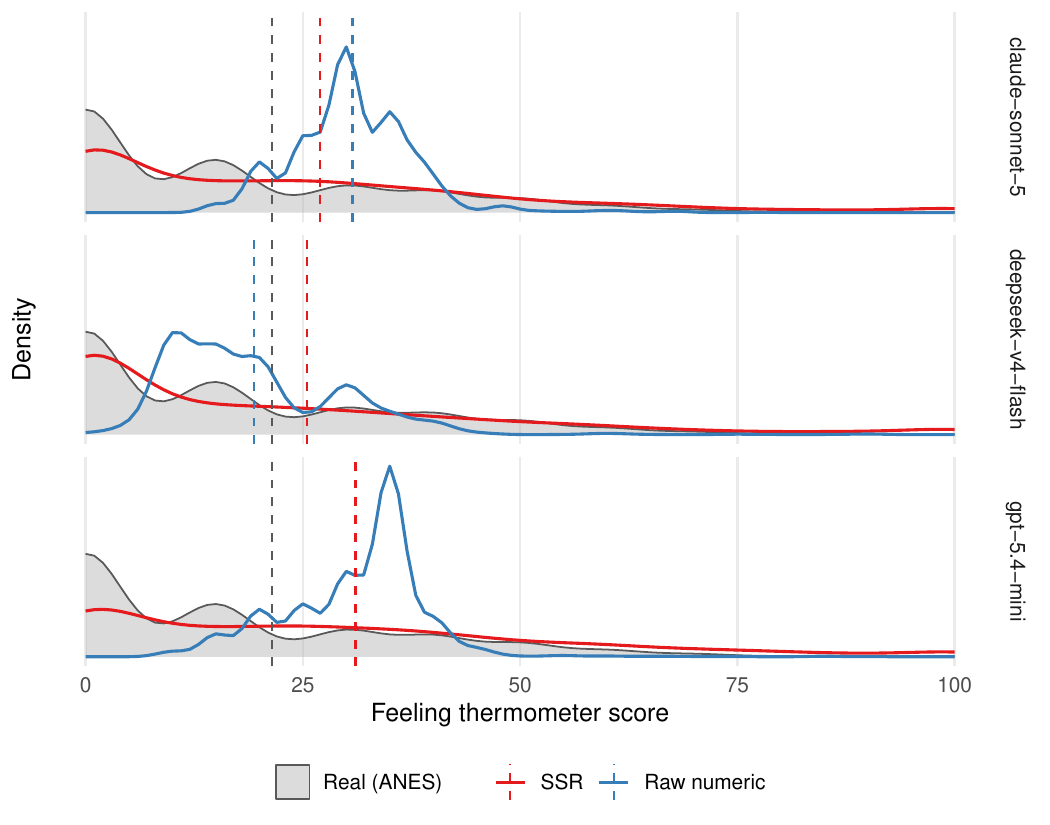}}

}

\caption{\label{fig-single-cell}Real and synthetic thermometer response
distributions for white Republicans asked about the Democratic Party.
The gray filled area is the real ANES 2016 distribution. Lines are
synthetic distributions from each model, generated by direct numeric
prompting (blue) or SSR (red). Dashed lines indicate the mean of each
distribution.}

\end{figure}%

\section{Results}\label{sec-results}

Figure~\ref{fig-single-cell} shows the real and synthetic response
distributions of white republican respondents asked about the Democratic
Party. Across all three models, we find that the raw numeric output is
peaked, with lower variance than the real distribution. In contrast, SSR
generated a distribution that is much closer to the real distribution,
with more variance and less peakedness. The means of the synthetic
distributions are similar to each other, and are also similar to the
mean of the real distribution. This example illustrates the general
trend in silicon sampling: while the means of raw numeric data
distributions are often close to the real means, the overall shape of
them are overly peaked, and the variance of the real distribution is not
captured. In contrast, SSR generates distributions that are much closer
to the real distributions, with more variance and less peakedness.

\begin{figure}

\centering{

\pandocbounded{\includegraphics[keepaspectratio]{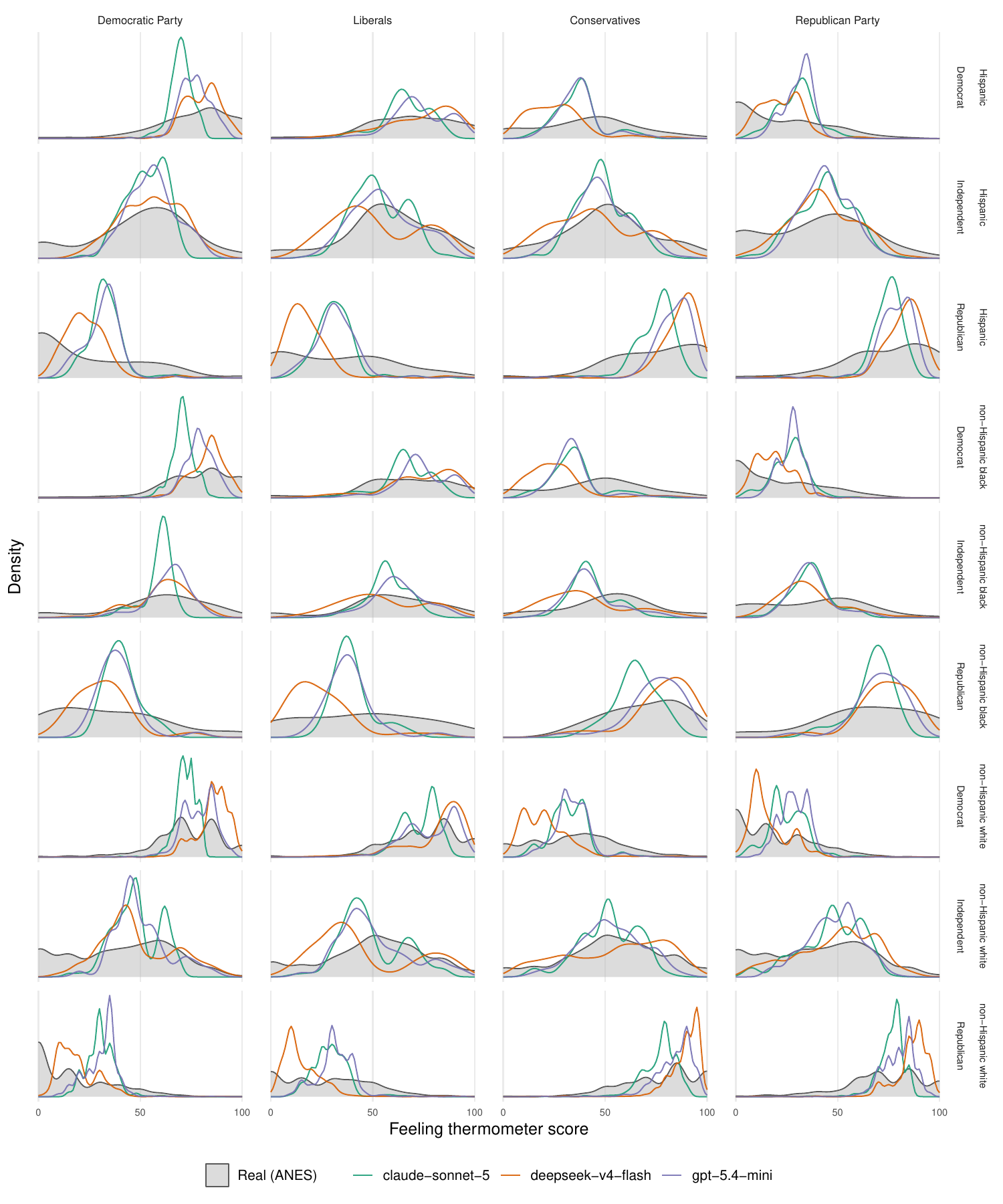}}

}

\caption{\label{fig-numeric-all-cells}Real and synthetic thermometer
response distributions under direct numeric prompting, by respondent
group and target group. Gray areas are actual ANES 2016 responses and
coloured lines are the three models.}

\end{figure}%

\begin{figure}

\centering{

\pandocbounded{\includegraphics[keepaspectratio]{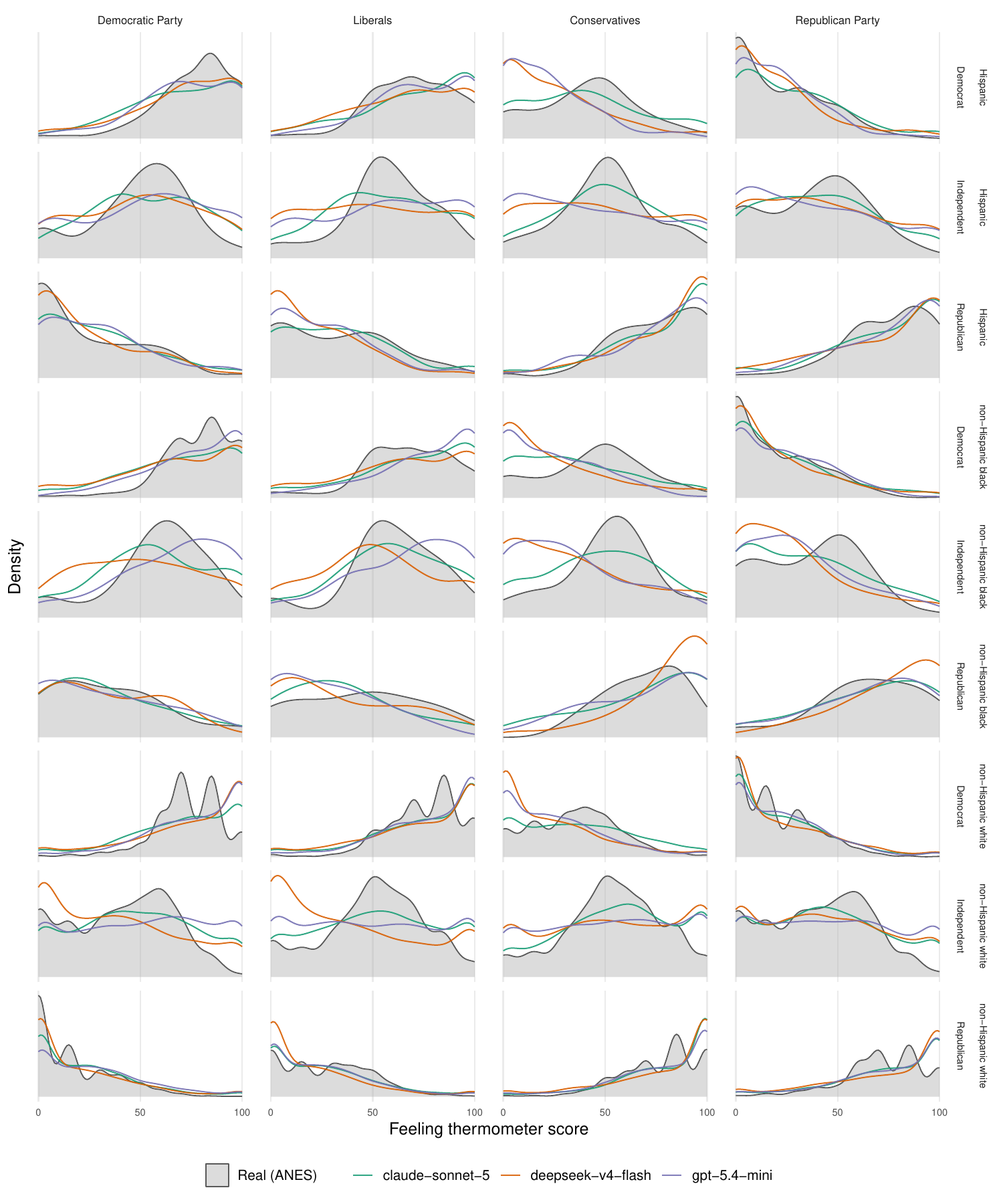}}

}

\caption{\label{fig-ssr-all-cells}Real and synthetic thermometer
response distributions under SSR, by respondent group and target group.
Gray areas are real ANES 2016 responses. Coloured lines are the three
models.}

\end{figure}%

Figure~\ref{fig-numeric-all-cells} and Figure~\ref{fig-ssr-all-cells}
expand on the example in Figure~\ref{fig-single-cell}, comparing the
synthetic distributions across all 36 respondent-target group pairs. We
see that results are consistent across all respondent groups and target
groups. Numeric responses display mode collapse and unrealistic
variance, consistent with the findings of Bisbee et al. (2024), which
highlights the continued issue of low variance in silicon sampling.
Newer models appear to not have meaningfully improved in this regard. We
see that SSR distributions match the shape of the real distributions
much more closely, with \texttt{claude-sonnet-5} performing the best.

\begin{figure}

\centering{

\pandocbounded{\includegraphics[keepaspectratio]{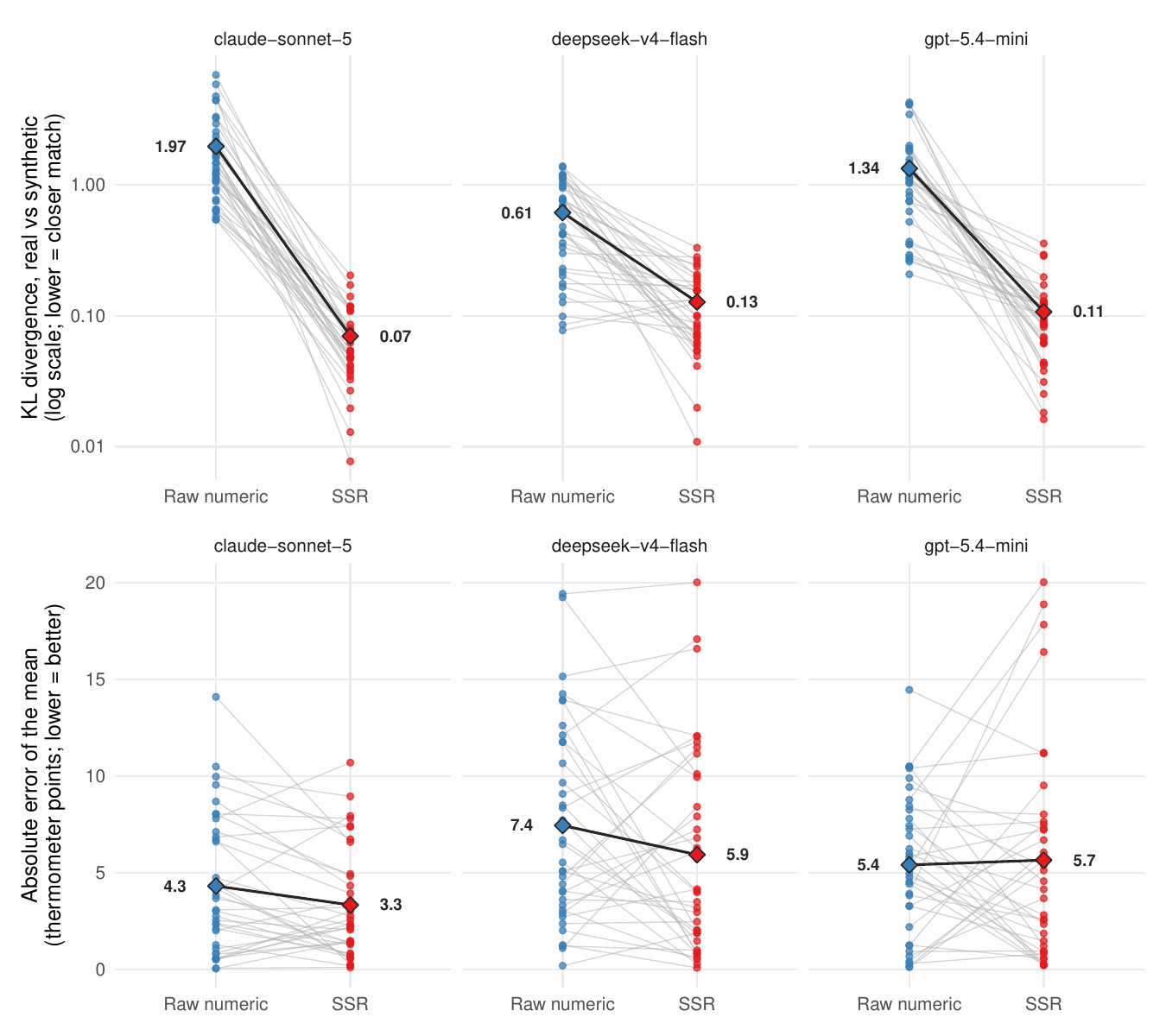}}

}

\caption{\label{fig-fit-metrics}Change in fit when moving from raw
numeric prompting to SSR. Each grey line is one respondent-group x
target cell (36 per model). The diamonds and printed values are means
over cells. The top panel is the Kullback--Leibler divergence between
synthetic and real response distributions. The bottom panel is the
absolute error of the synthetic mean.}

\end{figure}%

\begin{figure}

\centering{

\pandocbounded{\includegraphics[keepaspectratio]{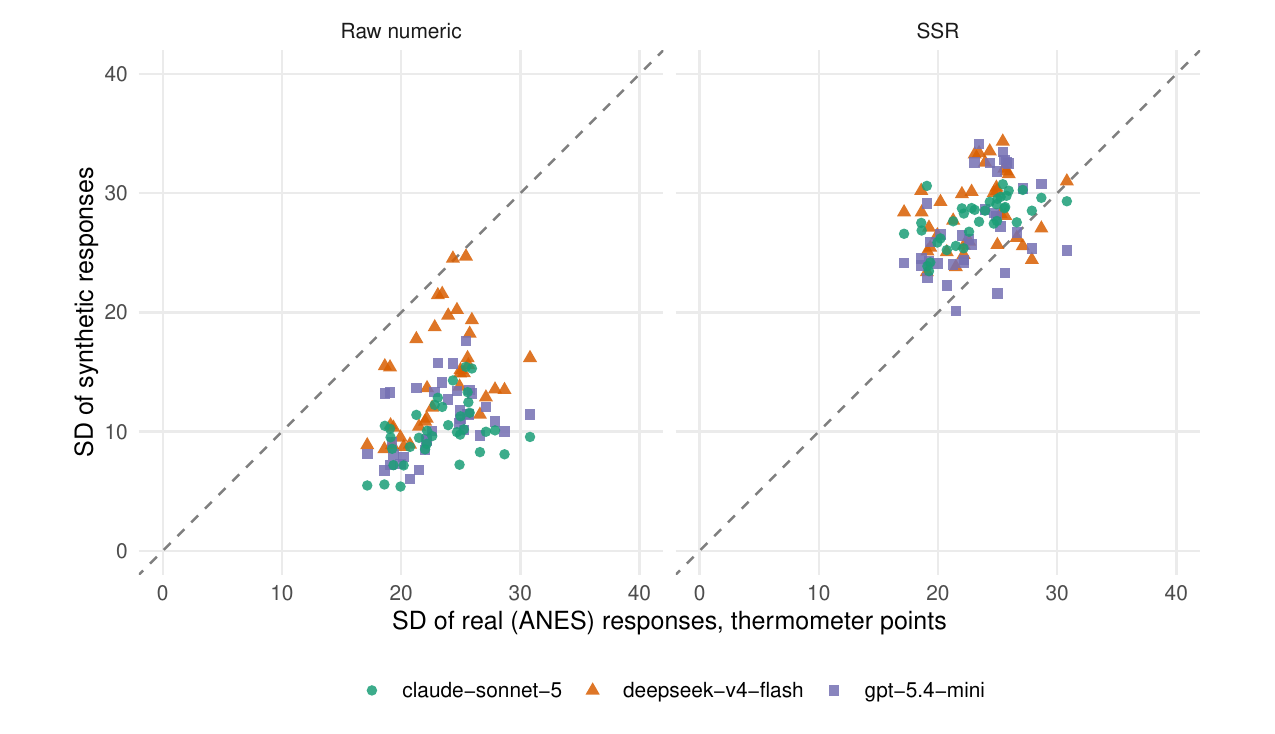}}

}

\caption{\label{fig-sd-calibration}Variance calibration of silicon
samples. Each point compares the standard deviation of real ANES
responses in one respondent-group x target cell (x-axis) with the
standard deviation of synthetic responses for the same cell (y-axis).
The left panel is the raw numeric output. The right panel is SSR.}

\end{figure}%

Beyond visual inspection, we quantified the improvements in the fidelity
of SSR distributions over raw numeric outputs.
Figure~\ref{fig-fit-metrics} shows substantial improvements in KL
divergence across all models. All models improved in KL divergence when
moving from raw numeric output to SSR, with \texttt{claude-sonnet-5}
performing the best. Meanwhile, mean absolute error of the synthetic
means remains largely unchanged when moving from raw numeric output to
SSR, with only a slight increase in error for \texttt{gpt-5.4-mini}.
This suggests that SSR improves the shape of synthetic distributions
without harming the accuracy of the synthetic means.
Figure~\ref{fig-sd-calibration} show that the standard deviation of
synthetic responses matches the standard deviation of real responses
much more closely under SSR than raw numeric output, further supporting
the conclusion that SSR improves the variance of synthetic
distributions.

We briefly acknowedge the concern that we fit a temperature parameter to
minimize KL divergence, then quantify improvements using KL divergence.
However, we note a couple of points. First, we only fit a single global
temperature parameter, which is applied to all respondent groups and
target groups. Hence, we are not overfitting to any particular group or
question. Another potential concern is that the temperature parameter in
the SSR process learns to simply flatten individual response
distributions, which would reduce KL divergence over the mode collpased
numeric distributions, but not truly improve the fidelity of the
synthetic distributions. To this point we would note that the fit
temperature parameter \(\mathcal{T}\) was 0.2, which is quite low, and
actually generates individual distributions that are quite peaked. We
found that KL divergence grew monotonely as \(\mathcal{T}\) increased,
implying that the SSR process is not simply flattening distributions to
uniform. The variance is coming from more varied text responses.

\subsection{Results on 2020 ANES Test
Set}\label{results-on-2020-anes-test-set}

To ensure that our temperature parameter and anchor choices were not
overfit to the 2016 ANES data, we applied the same parameter,
\(\mathcal{T} = 0.2\), and anchors to the 2020 ANES dataset, and
analyzed the response distributions produced by SSR. After cleaning the
raw 2020 ANES data using the same procedure as Bisbee et al. (2024) we
had 6630 unique personas. We made no changes to the prompting or the SSR
process.

\begin{figure}

\centering{

\pandocbounded{\includegraphics[keepaspectratio]{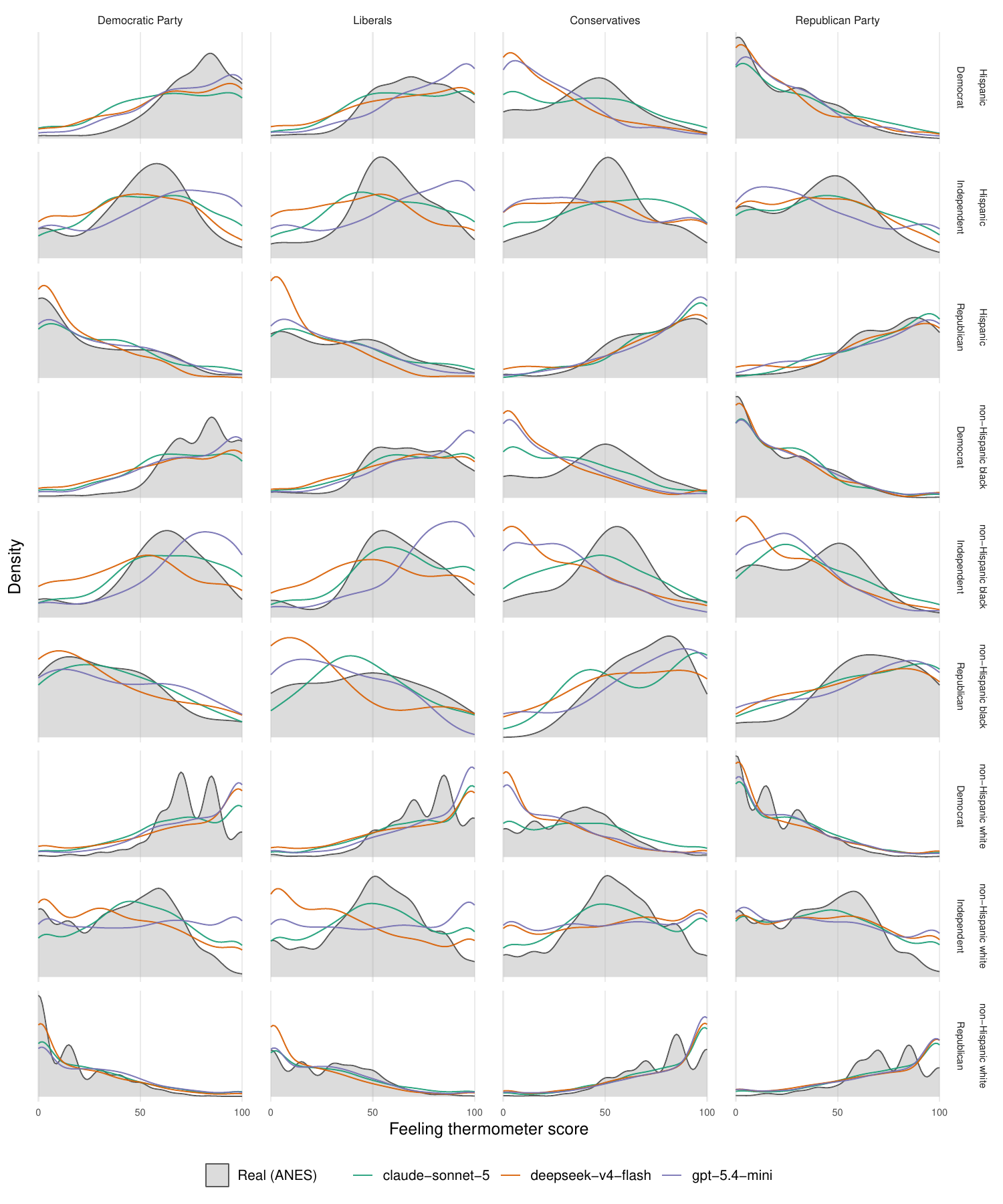}}

}

\caption{\label{fig-ssr-all-cells-2020}Real and synthetic 2020
thermometer response distributions, by respondent group and target
group. Gray areas are real ANES 2020 responses. Coloured lines are SSR
responses from all models using the global temperature parameter learned
on 2016 data.}

\end{figure}%

\begin{table}

\centering{

\fontsize{12.0pt}{14.0pt}\selectfont
\begin{tabular*}{1\linewidth}{@{\extracolsep{\fill}}lrrr}
\toprule
{\bfseries Calibration Method} & {\bfseries DeepSeek} & {\bfseries GPT} & {\bfseries Claude} \\ 
\midrule\addlinespace[2.5pt]
Raw Numeric 2016 & 0.61 & 1.34 & 1.97 \\ 
SSR 2016 with fit parameter & 0.13 & 0.11 & 0.07 \\ 
SSR 2020 with parameter fit to 2016 & 0.11 & 0.07 & 0.08 \\ 
\bottomrule
\end{tabular*}

}

\caption{\label{tbl-kl-summary}Summary of mean KL divergence between
real and synthetic distributions across all respondent groups and target
groups, for 2016 and 2020 ANES data. KL divergence is a measure of how
well the synthetic distribution approximates the real distribution, with
lower values indicating a better match.}

\end{table}%

Applying SSR with \(\mathcal{T}=0.2\) to 2020 ANES data, we obtained
very similar results to the 2016 ANES data (see
Figure~\ref{fig-ssr-all-cells-2020}). Table~\ref{tbl-kl-summary} shows
that the KL divergence between the real and SSR synthetic distributions
is very similar in 2020 as it was in 2016 when using the same global
temperature parameter. This suggests that the global temperature
parameter learned from the 2016 ANES data generalizes well to the 2020
ANES data.

\section{Discussion}\label{sec-discussion}

Overall, we find that applying SSR with only one global temperature
parameter leads to substantial improvements in the fidelity of synthetic
response distributions. The issue of unrealistically low variance is
effectively mitigated by calibrating a single temperature parameter,
without harming the mean absolute error of the synthetic distributions.
The overall shape of the response distributions are also improved, as
seen in Figure~\ref{fig-fit-metrics}, with the KL divergence between the
real and synthetic distributions dropping substantially in all models
when applying SSR over raw numeric output. Using the 2020 ANES data as a
test set, we find that the global temperature parameter learned from the
2016 ANES data generalizes well to the 2020 ANES data, lying very close
to the optimal global temperature for the 2020 set.

Despite a substantial improvement in KL divergence, the SSR process does
not correct some systematic biases in the synthetic response
distributions. As noted by Bisbee et al. (2024), the LLMs tend to
produce more extreme responses than are present in the real data, and
this bias is not corrected by SSR. In both the raw numeric output as
well as in the SSR output, the ratings of synthetic Democrats on
Conservatives are noticeably lower than the ratings of real Democrats.
So, while SSR improves the overall shape of the synthetic response
distributions, we see that biases of the model are not corrected by this
process. This suggests that this bias towards extreme opinions is not
due to an issue of LLMs having difficulty generating numbers, but rather
a more fundamental issue with the LLMs themselves and their
understanding of politics and psychology.

Overall, we find that SSR is an effective method for improving the
fidelity of synthetic response distributions on political surveys,
primarily improving on the issue of low variance in LLM responses. This
simple post-processing step can be applied to thermometer-style
questions, and requires only a small calibration set of real data to
learn temperature.

LLMs are increasingly being used by the public to inform their political
opinions. The New York Times (2026b) finds that Americans are
increasingly turning to LLMs for researching political issues, in large
enough numbers that some campaign strategists are prioritizing posting
material online in formats optimized for LLMs to read. It is therefore
important to understand the biases and limitations of LLMs in their
understanding of politics. We find that SSR does not correct for
systematic biases in the LLMs responses, and further research is needed
to understand the underlying causes of these biases and how to mitigate
them.

\newpage

\section*{References}\label{references}
\addcontentsline{toc}{section}{References}

\protect\phantomsection\label{refs}

\newpage

\appendix

\section{Appendix}\label{sec-appendix}

\subsection{Prompt Formats (identical to Bisbee et al.
(2024)):}\label{sec-prompts}

\subsubsection{System Prompt:}\label{system-prompt}

``It is {[}\texttt{YEAR}{]}. You are a {[}\texttt{AGE}{]} year-old,
{[}\texttt{MARST}{]}, {[}\texttt{RACETH}{]} {[}\texttt{GENDER}{]} with
{[}\texttt{EDUCATION}{]} making {[}\texttt{INCOME}{]} per year, living
in the United States. You are {[}\texttt{IDEO}{]}, {[}\texttt{REGIS}{]}
{[}\texttt{PID}{]} who {[}\texttt{INTEREST}{]} pays attention to what's
going on in government and politics.''

Where the variables are inputted as follows:

\begin{itemize}
\tightlist
\item
  {[}\texttt{YEAR}{]}: 2016 or 2020
\item
  {[}\texttt{AGE}{]}: age in years of ANES respondent
\item
  {[}\texttt{RACETH}{]}: non-Hispanic white, non-Hispanic black, or
  Hispanic
\item
  {[}\texttt{GENDER}{]}: male or female
\item
  {[}\texttt{MARST}{]}: divorced, married, separated, single, or widowed
\item
  {[}\texttt{EDUCATION}{]}: a high school diploma, some college but no
  degree, a bachelor's degree or more
\item
  {[}\texttt{INCOME}{]}: \$30k, \$50k, \$80k, \$100k, \$150k or more
\item
  {[}\texttt{IDEO}{]}: an extremely liberal, a liberal, a slightly
  liberal, a moderate, a slightly conservative, a conservative, an
  extremely conservative
\item
  {[}\texttt{REGIS}{]}: registered, unregistered
\item
  {[}\texttt{PID}{]}: Democrat, Independent, Republican
\item
  {[}\texttt{INTEREST}{]}: never, sometimes, frequently, regularly,
  always
\end{itemize}

\subsubsection{User Prompt}\label{user-prompt}

\paragraph{User Prompt for numeric
output:}\label{user-prompt-for-numeric-output}

``Provide responses from this person's perspective. Use only knowledge
about politics that they would have. Format the output as a tsv table
with the following format: group,thermometer,explanation,confidence

The following questions ask about individuals' feelings toward different
groups. Responses should be given on a scale from 0 (meaning cold
feelings) to 100 (meaning warm feelings). Ratings between 50 degrees and
100 degrees mean that you feel favorable and warm toward the group.
Ratings between 0 degrees and 50 degrees mean that you don't feel
favorable toward the group and that you don't care too much for that
group. You would rate the group at the 50 degree mark if you don't feel
particularly warm or cold toward the group.

How do you feel toward the following groups?

The Democratic Party? The Republican Party? Liberals? Conservatives?
``\,''

\paragraph{User Prompt for text
output:}\label{user-prompt-for-text-output}

``Provide responses from this person's perspective. Use only knowledge
about politics that they would have. Format the output as a tsv table
with the following format: group,opinion,explanation

The following questions ask about individuals' feelings toward different
groups. Responses should be your genuine feelings toward each group, not
what you think is socially desirable. Please elaborate in your answers.

How do you feel toward the following groups?

The Democratic Party? The Republican Party? Liberals? Conservatives?

\subsection{Anchor Points for Semantic Similarity Rating
(SSR):}\label{sec-anchor-points}

To implement SSR, we defined anchor points on the thermometer scale at
0, 25, 50, 75, and 100, with the following corresponding text
descriptions:

Anchor 0 = ``I have a very unfavorable opinion of {[}GROUP{]}. I
strongly dislike {[}GROUP{]} and reject almost everything {[}GROUP{]}
stands for.'' Anchor 25 = ``On balance I lean unfavorable toward
{[}GROUP{]}. There is more about {[}GROUP{]} that I disagree with than
agree with.'' Anchor 50 = ``I don't have a strong opinion of {[}GROUP{]}
either way. I see about as much to like as to dislike about {[}GROUP{]},
and my feelings are genuinely mixed.'' Anchor 75 = ``On balance I lean
favorable toward {[}GROUP{]}. There is more about {[}GROUP{]} that I
agree with than disagree with.'' Anchor 100 = ``I have a very favorable
opinion of {[}GROUP{]}. I strongly like {[}GROUP{]} and support almost
everything {[}GROUP{]} stands for.''

\end{document}